\documentstyle[epsfig,axodraw,12pt]{elsart}
\input epsf
\newcommand{\be}[1]{\begin{equation} \label{(#1)}}
\newcommand{\ee}{\end{equation}}
\newcommand{\ba}[1]{\begin{eqnarray} \label{(#1)}}
\newcommand{\ea}{\end{eqnarray}}
\newcommand{\nn}{\nonumber}
\newcommand{\rf}[1]{(\ref{(#1)})}
\def\fig#1{{Fig. (\ref{#1})}}

\def\rp{$R_p \hspace{-1em}/\;\:$ }

\def\rpm{R_p \hspace{-0.8em}/\;\:}

\def\nm{\hbox{$\nu_\mu$ }}
\def\nt{\hbox{$\nu_\tau$ }}

\def\21{$SU(2) \otimes U(1) $}
\def\npb#1#2#3{{\it Nucl.\ Phys.\ }{\bf B #1} (#2) #3}
\def\plb#1#2#3{{\it Phys.\ Lett.\ }{\bf B #1} (#2) #3}
\def\prd#1#2#3{{\it Phys.\ Rev.\ }{\bf D #1} (#2) #3}

\def\hepph#1{{\tt hep-ph/#1}}

\begin{document}

\begin{flushright}hep-ph/0009066\\ IFIC/00-048 \end{flushright}  

\begin{frontmatter}
  
  \title{Reconciling neutrino anomalies in a simple
    four-neutrino scheme with  R-parity violation}
  
  \author{M. Hirsch\thanksref{mahirsch}} \author{and}
  \author{J.~W.~F. Valle\thanksref{valle}} \address{Instituto de
    F\'{\i}sica Corpuscular -- C.S.I.C. - Universitat of Val\`encia,
    \\ Edificio Institutos de Paterna, Apartado de Correos 2085 \\ 
    46071 Val\`encia, Spain}

\thanks[mahirsch]{(a) mahirsch@neutrinos.uv.es}
\thanks[valle]{ (a) valle@neutrinos.uv.es}

\begin{abstract}
  
  We propose a simple extension of the MSSM based on extra compact
  dimensions which includes an \21 singlet superfield.  The fermion
  present in this superfield is the sterile neutrino, which combines
  with one linear combination of $\nu_e-\nu_{\mu}-\nu_{\tau}$ to form
  a Dirac pair whose mass accounts for the LSND anomaly. Its small
  mass can be ascribed to a volume suppression factor associated with
  extra compact dimensions. On the other hand the sterile neutrino 
  scalar partner can trigger the spontaneous violation of R-parity,
  thereby inducing the necessary mass splittings to fit also the solar
  and atmospheric neutrino data.  Thus the model can explain all
  neutrino oscillation data. It leads to four predictions for the
  neutrino oscillation parameters and implies that the atmospheric
  neutrino problem must include at least some $\nu_{\mu} \to \nu_s$
  oscillations, which will be testable in the near future. Moreover it
  also predicts that the lightest supersymmetric particle (LSP) decays
  visibly via lepton number violating modes, which could be searched
  for at present and future accelerators.

\end{abstract}
\end{frontmatter}

\section{Introduction}

It is well-known that within the standard framework with only three
light neutrinos it is impossible to reconcile solar and atmospheric
neutrino data ~\cite{ichepsolat00} with those of LSND~\cite{LSND}. A
solution to the atmospheric neutrino problem requires a $\Delta m^2$
of the order of $\Delta m^2 \simeq ({\rm few}) 10^{-3}$
$eV^2$~\cite{update00,Fornengo:2000sr}, while solar neutrino data
could be explained either by $\Delta m^2 \sim {\cal O}(10^{-5})$
$eV^2$ or $\Delta m^2 \sim {\cal O}(10^{-7})$
$eV^2$~\cite{update00,MSW00} (LMA or LOW solutions of the solar
neutrino problem), none of which can be reconciled with the scale
$\Delta m^2 \sim 0.3 - 1$ $eV^2$ indicated by the LSND
experiment~\cite{LSND}.  Since the LSND experiment has not been
confirmed independently, many theoretical papers have chosen to ignore
this result~\footnote{ The KARMEN experiment \cite{Eitel:2000ry} does
  rule out some parts of the parameter space favoured by LSND.
  However, it does not disprove LSND.}. 

A simple way to account for all neutrino data would be to postulate
the existence of an additional neutrino
state~\cite{Schechter:1980bn,Schechter:1980gr}, which due to the
constraints from LEP has to be mainly sterile. 

\begin{figure}
  \setlength{\unitlength}{1mm}
\begin{picture}(0,100)
\put(30,-70)
{\mbox{\epsfig{figure=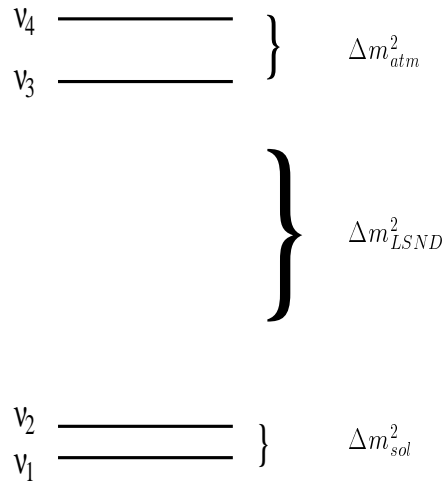,height=18.0cm,width=12.0cm}}}
\end{picture}

\caption[]{Four-neutrino spectrum fitting solar, atmospheric and LSND data. }
\label{4nuspec}
\end{figure}

However, even including a sterile neutrino, there are essentially only
two neutrino spectra allowed by the data. One of the possibilities is
shown in \fig{4nuspec}. It consists of two nearly degenerate pairs of
neutrinos separated by a gap of the order of the LSND scale. The two
pairs then have to be split by $\Delta m^2_{ij}$ which correctly fit
solar and atmospheric data. For recent analyses of atmospheric and
solar neutrino data in four-neutrino models see
ref.~\cite{Yasuda:2000de,Giunti:2000wt}.  In \fig{4nuspec} we show
only the case where the lower $\Delta m^2_{ij}$ corresponds to the
solar scale. A second spectrum with $\Delta m^2_{atm} \leftrightarrow
\Delta m^2_{sol}$ is equally well allowed by the data \footnote{The
  case where $\Delta m^2_{sol}$ and $\Delta m^2_{atm}$ are exchanged
  is allowed, but can not be realized in our model.}.

The question then is: Can one make sense of such a neutrino spectrum
theoretically? 

A number of attempts can be found in the
literature~\cite{Valle:2000yr}.  While originally motivated by the
desire to account for hot dark matter
~\cite{Peltoniemi:1993ss,Peltoniemi:1993ec,Caldwell:1993kn} soon after
LSND results came into existence, it was realized that schemes leading
to these spectra would easily fit LSND results together with solar and
atmospheric data.

Here we propose, what we believe to be one of the simplest particle
physics model for the inclusion of a sterile neutrino state into the
spectrum. It is based on a minimal extension of the MSSM with one
additional singlet superfield field.  The fermionic component of the
singlet combines with (one of) the active neutrinos to form a light
Dirac state at the LSND scale. Its scalar neutrino component develops
a nonzero vacuum expectation value (vev)~\cite{Ross:1985yg,SBRpV},
breaking R-parity spontaneously~\cite{rpv99} and effectively
generating bilinear R-parity violating terms in the
superpotential~\cite{epsrad,epsothers}.

Supersymmetry with bilinear R-parity violation has been shown to
provide a predictive theory for solar and atmospheric neutrino
oscillations in which the neutrino masses and mixing angles are all
determined in terms of the three fundamental bilinear
terms~\cite{Hirsch:2000ef,rphen00}. In the context of the present
4-neutrino scheme the breaking of R-parity leads to Majorana masses
for the neutrinos, splitting the Dirac neutrino into a quasi-Dirac
pair~\cite{Valle:1983yw}, and giving mass to one additional neutrino
state. It is this breaking of R-parity that leads to the solar and
atmospheric oscillations. The model has therefore, despite being
minimalistic, all the basic ingredients for solving solar and
atmospheric neutrinos in addition to the LSND data.  Moreover it
extends the predictivity of the bilinear \rp model to the 4-neutrino
case: four parameters are predictable, out of a total of ten (4 masses
and six mixing angles)\footnote{We will assume CP invariance
  throughout this paper.}.  In contrast with the early models, here
the smallness of the LSND scale arises without appealing to a
radiative mechanism, but due to the volume factor associated with the
canonical normalization of the wave-function of the bulk field in the
compactified dimensions, as suggested in
ref.~\cite{Ioannisian:1999sw}.

This paper is organized as follows. In the next section we describe
the main features of the model. Then, we discuss some numerical
results, exploring what are the parameter ranges which could fit the
neutrino data. We also discuss, how this model can be tested by both,
future neutrino measurements as well as at accelerators, before giving
a short conclusion.

\section{The model}

The model we advocate can be regarded  as a simple extension of
the bilinear R-parity broken MSSM ~\cite{epsrad,epsothers}.  The
minimal extension beyond the MSSM particle content is to include a
single right-handed neutrino.  Thus to the superpotential of the MSSM
we add,
\be{spot}
W = W_{MSSM} + h^{\nu}_i {\widehat L_i} {\widehat H_u} 
                         {\widehat \nu^c}.
\ee

The last term in the equation above gives three Dirac mass terms
\be{defdi}
d_i = h^{\nu}_i v_u
\ee
giving automatically a Dirac mass to one linear combination of
$\nu_e-\nu_{\mu}-\nu_{\tau}$ once $ H_u$ develops a nonzero vev. The
size of this entry is governed by the magnitude of the corresponding
Yukawa couplings $h^{\nu}_i$.

However, within a supersymmetric framework the $h^{\nu}_i {\widehat
  L_i} {\widehat H_u} {\widehat \nu^c}$ term can have also other
consequences.

In SUSY models the scalar partner in the $\nu^c$ superfield can,
depending on the parameters of the model, acquire a vacuum expectation
value ($v_R$) through the usual Higgs mechanism, breaking R-parity
spontaneously, as proposed in ref.~\cite{SBRpV}. For this all we need
is to have its squared mass negative at the weak scale, like in the
standard model Higgs mechanism. First note that this may happen as a
result of the primordial choice of non-universal boundary conditions
at the unification scale which make $m^2_{\nu^c}$ different from the
\21 non-singlet soft scalar masses. Alternatively it might, under some
circumstances, be induced dynamically via radiative
corrections~\cite{SBRpVrad}.

Such a right-handed sneutrino vev, however, produces effectively
bilinear terms in the superpotential via
\be{defeps}
\epsilon_i = h^{\nu}_i v_R
\ee
It is well-known that the existence of such lepton number violating 
(R-parity breaking) bilinears lead automatically to Majorana masses 
for the neutrinos. 

In the following we will assume a non-zero $v_R$. With this one can
immediately write down the mass matrix of the model. Choosing as basis
${\Psi'_{0}}^T = (\nu_e, \nu_{\mu}, \nu_{\tau},\nu^c, -i\lambda', -i
\lambda_3, \psi_{H_d}^1,\psi_{H_u}^2)$ we find that the
neutrino-neutralino mass matrix of this model is given by:
\be{mas8}
{\cal M}_0^8 =  \left(
                    \begin{array}{cc}
                    m_D & m_{\rpm} \\
                    m_{\rpm}^T & {\cal M}_{\chi^0} \\
                    \end{array}
              \right).
\ee
where ${\cal M}_{\chi^0}$ is the MSSM neutralino mass matrix, 
\be{mssmmass}
{\cal M}_{\chi^0} =  \left(
                        \begin{array}{cccc}
 M_1 & 0   & -\frac{1}{2}g' v_d &  \frac{1}{2} g' v_u  \\
 0   & M_2 &  \frac{1}{2}g  v_d & -\frac{1}{2} g  v_u  \\
  -\frac{1}{2} g' v_d &  \frac{1}{2} g v_d & 0 & -\mu  \\
   \frac{1}{2} g' v_u & -\frac{1}{2} g v_u & -\mu & 0 \\
 \end{array}
                     \right).
\ee
and the Dirac mass matrix $m_D$ is given by,
\be{dmass}
m_D  =  \left(
                        \begin{array}{cccc}
        0 & 0 & 0 & \frac{1}{\sqrt{2}} h_1 v_u  \\
        0 & 0 & 0 & \frac{1}{\sqrt{2}} h_2 v_u  \\
        0 & 0 & 0 & \frac{1}{\sqrt{2}} h_3 v_u  \\
     \frac{1}{\sqrt{2}} h_1 v_u &  \frac{1}{\sqrt{2}} h_2 v_u &
     \frac{1}{\sqrt{2}} h_3 v_u & 0 \\
 \end{array}
                     \right).
\ee
Finally, $m_{\rpm}$ contains the entries induced by the breaking of
R-parity.  These can be read off from the mass matrix of the
spontaneous model as,
\be{rpmass}
m_{\rpm} =   \left(
            \begin{array}{cccc}
  -\frac{1}{2}g'\langle {\tilde \nu}_{e} \rangle & 
   \frac{1}{2}g\langle {\tilde \nu}_{e} \rangle & 0 & -\epsilon_e \\
  -\frac{1}{2}g'\langle {\tilde \nu}_{\mu} \rangle & 
   \frac{1}{2}g\langle {\tilde \nu}_{\mu} \rangle & 0 & -\epsilon_{\mu} \\
  -\frac{1}{2}g'\langle {\tilde \nu}_{\tau} \rangle & 
   \frac{1}{2}g\langle {\tilde \nu}_{\tau} \rangle & 0 & -\epsilon_{\tau} \\
   0 & 0 & 0 & \sum h^{\nu}_i \langle {\tilde \nu}_{i} \rangle \\
                    \end{array}
              \right).
\ee
Here $\langle {\tilde \nu}_{i} \rangle$ are the left-handed sneutrino
vevs, which are in general non-zero due to the minimization conditions
of the scalar potential once we have non-zero $\epsilon_i$.  Note that
both $m_{\rpm}$ and $m_D$ have such a structure that with either term
present only one neutrino would gain a mass. Only if $m_D$ and
$m_{\rpm}$ are not completely ``aligned'' to each other (i.e. if at
least one of the ratios $h_i/h_j$ differs from $\Lambda_i/\Lambda_j$
where $\Lambda_i = \epsilon_i v_d + \mu \langle {\tilde \nu}_{i}
\rangle$) we have three non-zero masses plus one massless state in the
spectrum.

Depending on the relative size of $|\Lambda|^2/M_{SUSY}^4$ and 
$|h^{\nu}|$, where, 

\be{abslam}
|\Lambda| = \sqrt{\Lambda_e^2+\Lambda_{\mu}^2+\Lambda_{\tau}^2}
\ee
\be{absh}
|h^{\nu}| = \sqrt{(h^{\nu}_1)^2+(h^{\nu}_2)^2+(h^{\nu}_3)^2}
\ee
the heaviest state will be either a quasi-Dirac pair or a Majorana
neutrino. Possible parameter choices which fit the neutrino
oscillation data are discussed in the next section.

Before closing this section let us comment on the required smallness
of $h^{\nu}_i$. One way to avoid having to simply postulate as a
phenomenological assumption is to assume the existence of extra
compact dimensions, probed only by gravity and possibly gauge-singlet
fields, which can lower the fundamental scales to the weak scale (TeV)
\cite{extra}.  There have been many recent papers applying the idea of
extra dimensions to neutrino physics~\cite{extranu}. Using the Gauss
law one can write
\be{gauss}
M^2_{Pl}\simeq(R \ M_F)^n M_F^2 ,
\ee
where $R$ is the compactification radius of the additional dimensions
and $M_F$ is the fundamental Planck scale, which in these theories can
be low.  A recent attempt in this direction has been presented in
ref.~\cite{Ioannisian:1999sw}.  It postulates that $M_F \simeq 10$ TeV
and a 4+n dimensional theory with $n=6$ for which the corresponding
value of $R$ is $R \simeq 10^{-12}$ cm.  This way the smallness of
$h^{\nu}_i$ (and hence of the LSND mass scale) will follow from the
volume factor associated with the canonical normalization of the
wave-function of the bulk field in the compactified dimensions.  Our
fourth light neutrino $\nu_s$ ($s$ for sterile) is identified with the
zero mode of the Kaluza-Klein states.  To first approximation the
sterile neutrino combines with a combination of the active neutrinos
(mainly \nm with some \nt) in order to form a Dirac neutrino with mass
in the eV range leaving the other two neutrinos massless.  Thus we can
apply exactly the same construction in the present case. In other
words, the present model may be regarded as a variant of that in
ref.~\cite{Ioannisian:1999sw} in which neutrino mass splittings are
now due to the breaking of R-parity. As we discuss below this has
important phenomenological advantages.

\section{Numerical results}

\begin{figure}
\setlength{\unitlength}{1mm}
\begin{picture}(0,70)
\put(-25,-85)
{\mbox{\epsfig{figure=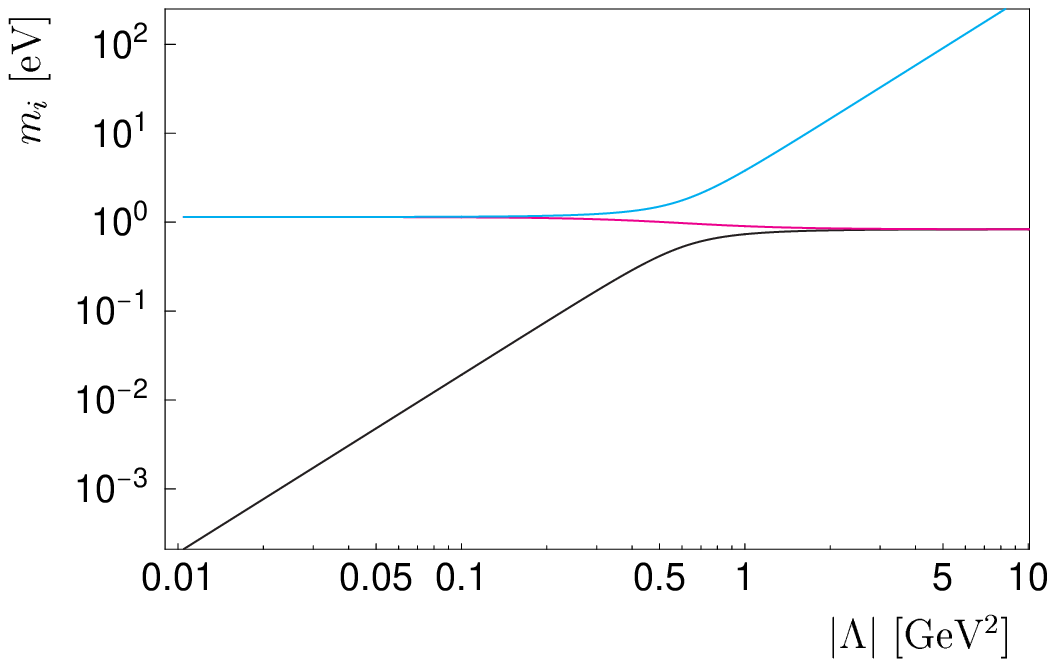,height=20.0cm,width=15.0cm}}}
\put(50,-85)
{\mbox{\epsfig{figure=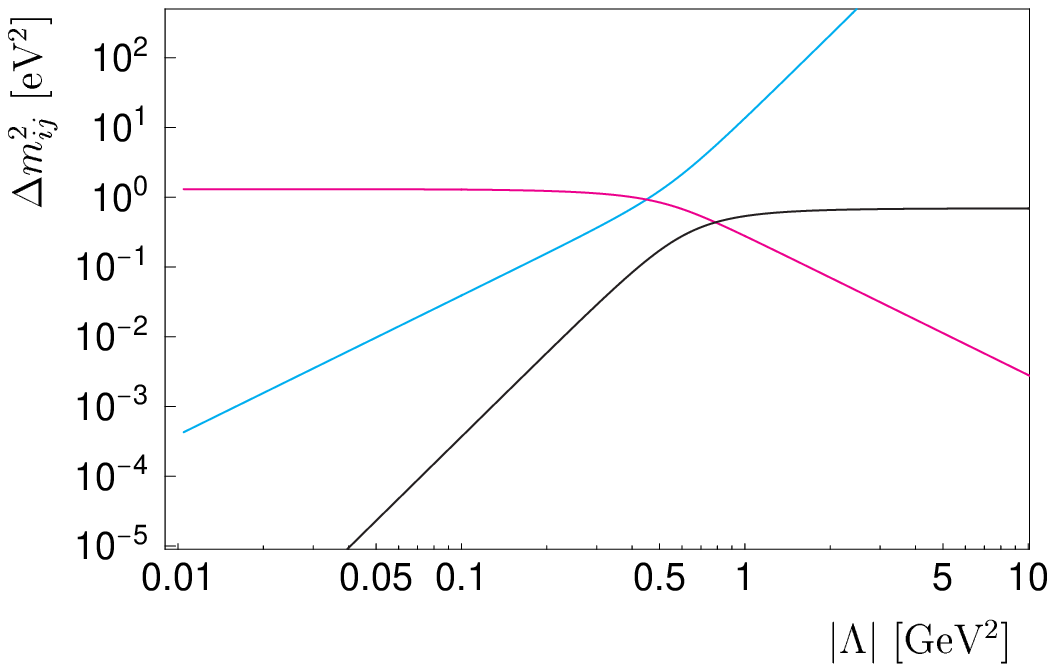,height=20.0cm,width=15.0cm}}}
\end{picture}
\caption[]{Example of calculated mass eigenvalues and corresponding mass 
squared differences as a function of $|\Lambda|$ for an arbitrary but 
fixed choice of other model parameters. To the left, mass eigenvalues. 
To the right, $\Delta m^2_{42}$, $\Delta m^2_{43}$ and $\Delta m^2_{21}$. 
Parameters which have been kept fixed 
for the calculation shown in this plot are $|h^{\nu}| = 5\times 10^{-12}$, 
$\mu = 500$ GeV, $M_2 = 200$ GeV and $\tan\beta =2.5$.}
\label{versuslam}
\end{figure}

We now turn to the pattern of neutrino masses and mixings arising from
our model. Starting with the masses we show in \fig{versuslam} the
eigenvalues which follow from diagonalization of the mass matrix for a
specific, though arbitrary choice of parameters as a function of
$|\Lambda|$. For small values of $|\Lambda|$ two neutrino states form
a Dirac pair.  Increasing the size of the R-parity breaking (while
keeping other parameters fixed) increases the mass splitting within
this pair, as well as the mass of the second mass eigenstate lying at
the solar neutrino scale. In this example, for $|\Lambda| \simeq 0.05$
$GeV^2$ the three mass squared differences are of the right order of
magnitude for solving the neutrino problems. Note, that the actual values 
of the MSSM parameters are not essential. Larger or smaller values of 
MSSM masses could be accounted for by a simple appropriate rescaling of 
$|\Lambda|$. 

One sees from \fig{versuslam} that for small values of $|\Lambda|$
$\nu_3$ and $\nu_4$ combine to form a quasi-Dirac pair. The splitting
between those two states increases with increasing $|\Lambda|$.

Turning now to mixings, a general weak interaction four-neutrino model
requires, in addition to the four masses, six mixing angles (and six
CP phases if CP is violated) in order to characterize the structure of
the weak leptonic current~\cite{Schechter:1980gr}.

As explained above the present model has a high degree of
predictivity, since now all 10 neutrino oscillation parameters are
given in terms of six independent quantities which may be conveniently
chosen as the $d_i$ and the alignment parameters $\Lambda_i$.  Due to
the small left-handed sneutrino vevs induced by the minimization of
the scalar potential the $\Lambda_i$ ratios are not proportional to 
the $h_i$ ratios, thereby breaking the projectivity of the neutrino 
mass matrix.  This is crucial to induce the solar neutrino oscillations.

Within any model producing a ($2,2$)-neutrino spectrum as shown in
\fig{4nuspec} the conversion probabilities relevant for reactor,
atmospheric and LSND experiments are modified with respect to the
usual 2-generation formula. \footnote{Also the solar neutrino survival
  probability will receive a modification. However, CHOOZ and
  atmospheric neutrino results imply that this correction is
  negligible.} In order to compare the experimental data with the
results of our model we therefore give in the following the conversion
(and survival) probabilities corresponding to the neutrino spectrum of
\fig{4nuspec}.

For the reactor neutrinos one finds after some trivial algebraic 
manipulations,
\ba{cpchooz}\nn
P_{\nu_e \to  \nu_e}^{CHOOZ} = 
1 &-& 4 \sin^2(\Delta m^2_{LSND} \frac{L}{E}) |(1 - U_{e3}^2-U_{e4}^2) 
(U_{e3}^2+U_{e4}^2) | \\
& -& 4 \sin^2(\Delta m^2_{atm} \frac{L}{E}) |U_{e3}|^2 |U_{e4}|^2
\ea

\begin{figure}
\setlength{\unitlength}{1mm}
\begin{picture}(0,70)
\put(-25,-85)
{\mbox{\epsfig{figure=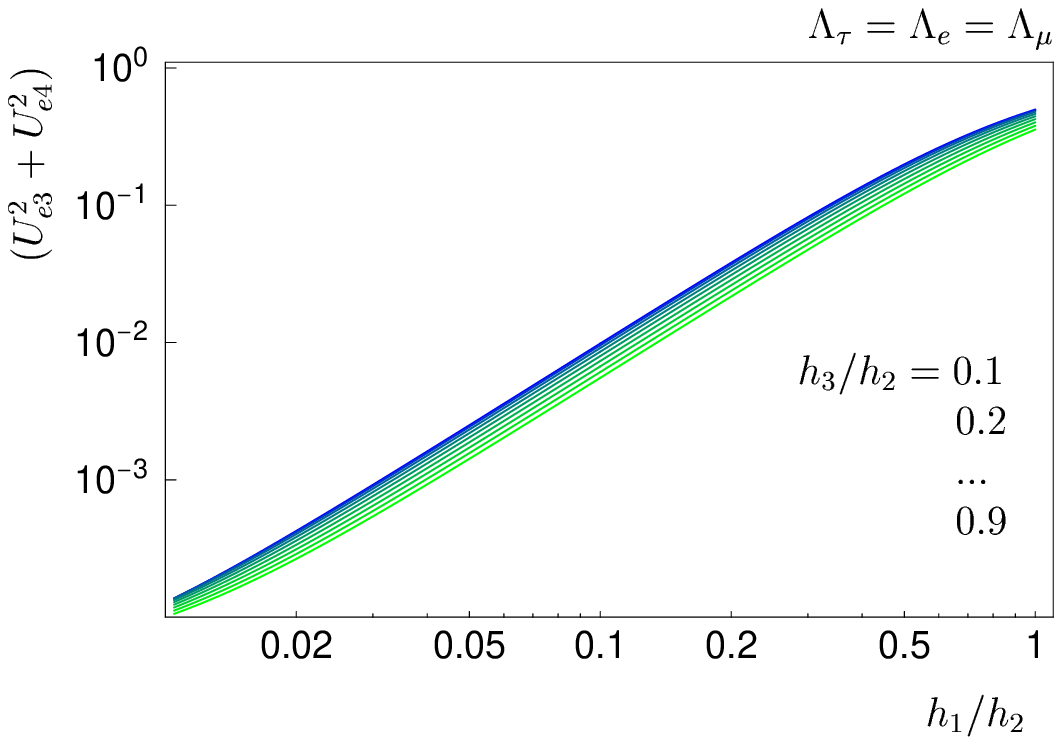,height=20.0cm,width=15.0cm}}}
\put(50,-85)
{\mbox{\epsfig{figure=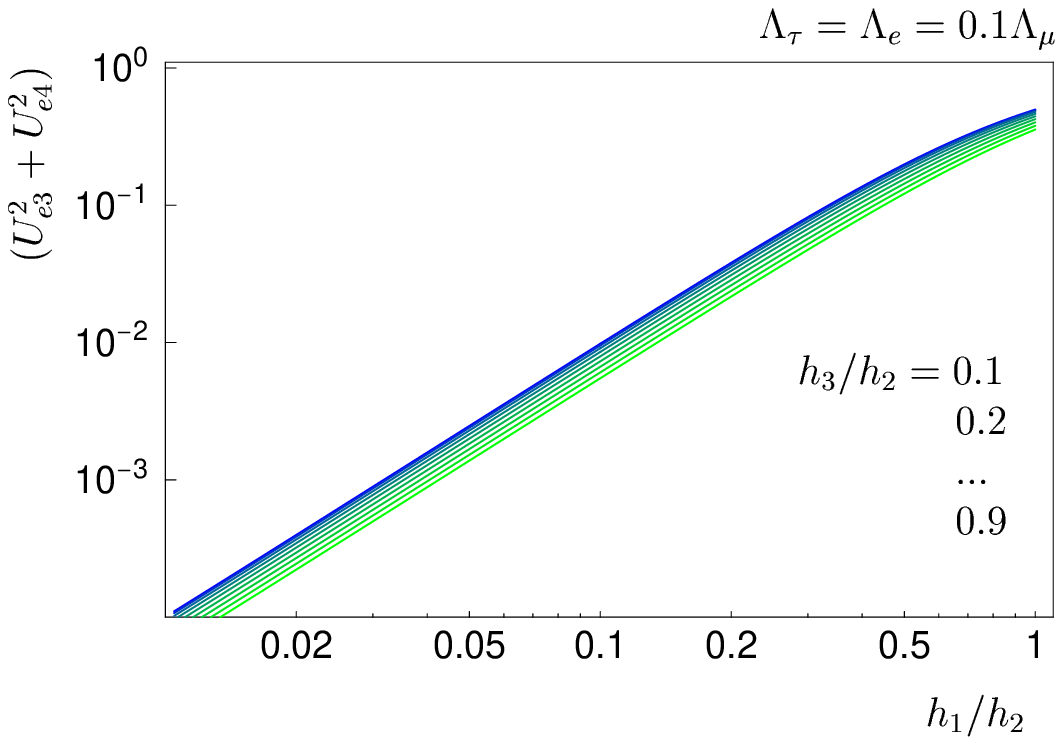,height=20.0cm,width=15.0cm}}}
\end{picture}
\caption[]{Two examples of ($U_{e3}^2+U_{e4}^2$) as a function of 
$h_1/h_2$ for different choices of $h_3/h_2$. This quantity is constrained 
by CHOOZ \cite{chooz} and implies $h_1 < h_2$. Only a rather weak 
dependence on other model parameters is found.}
\label{chooz}
\end{figure}

In \fig{chooz} we show two examples of ($U_{e3}^2+U_{e4}^2$) as a
function of $h_1/h_2$ for different choices of $h_3/h_2$. This
quantity is constrained by the non-observation of oscillations at 
reactor experiments and has to be rather small, implying that 
$h_1 < h_2$. 

Similarly for the LSND oscillation probability one finds,
\ba{cplsnd}\nn
P_{{\bar \nu}_{\mu} \to  {\bar \nu}_e}^{LSND} &=& 
 4 \sin^2(\Delta m^2_{LSND} \frac{L}{E}) |(U_{e1}U_{\mu 1}+U_{e2} U_{\mu 2}) 
(U_{e3} U_{\mu 3}+U_{e4} U_{\mu 4}) | \\
& =: &  \sin^2(\Delta m^2_{LSND} \frac{L}{E}) \sin^2(2\theta_{LSND,eff})
\ea

\begin{figure}
\setlength{\unitlength}{1mm}
\begin{picture}(0,70)
\put(-25,-85)
{\mbox{\epsfig{figure=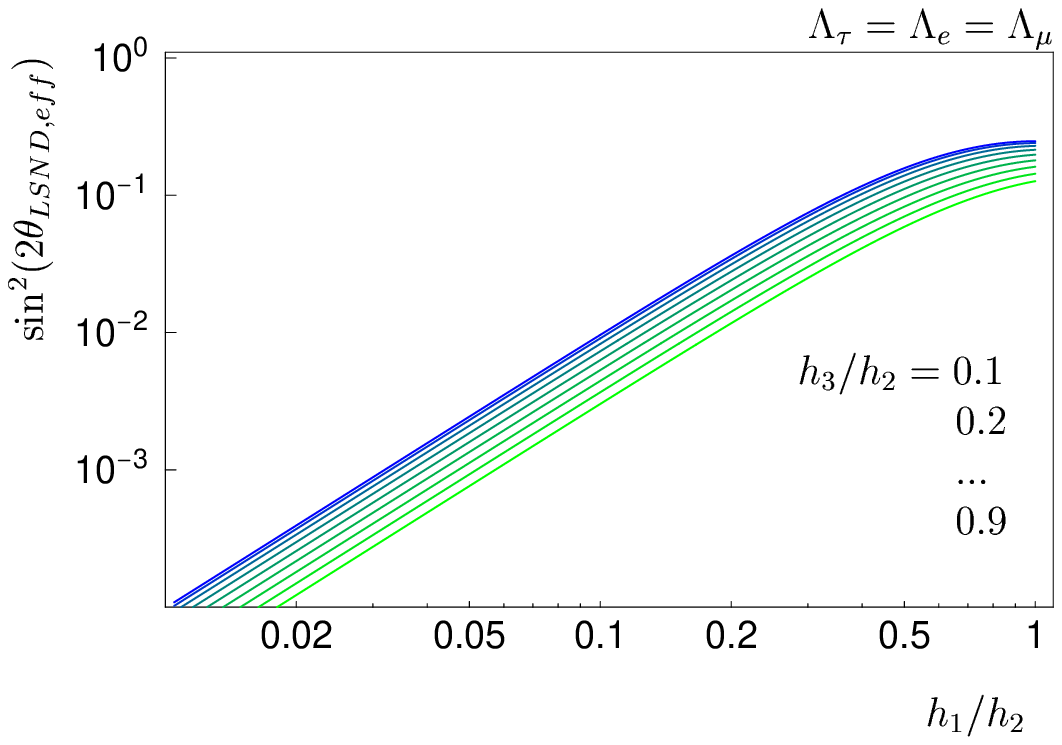,height=20.0cm,width=15.0cm}}}
\put(50,-85)
{\mbox{\epsfig{figure=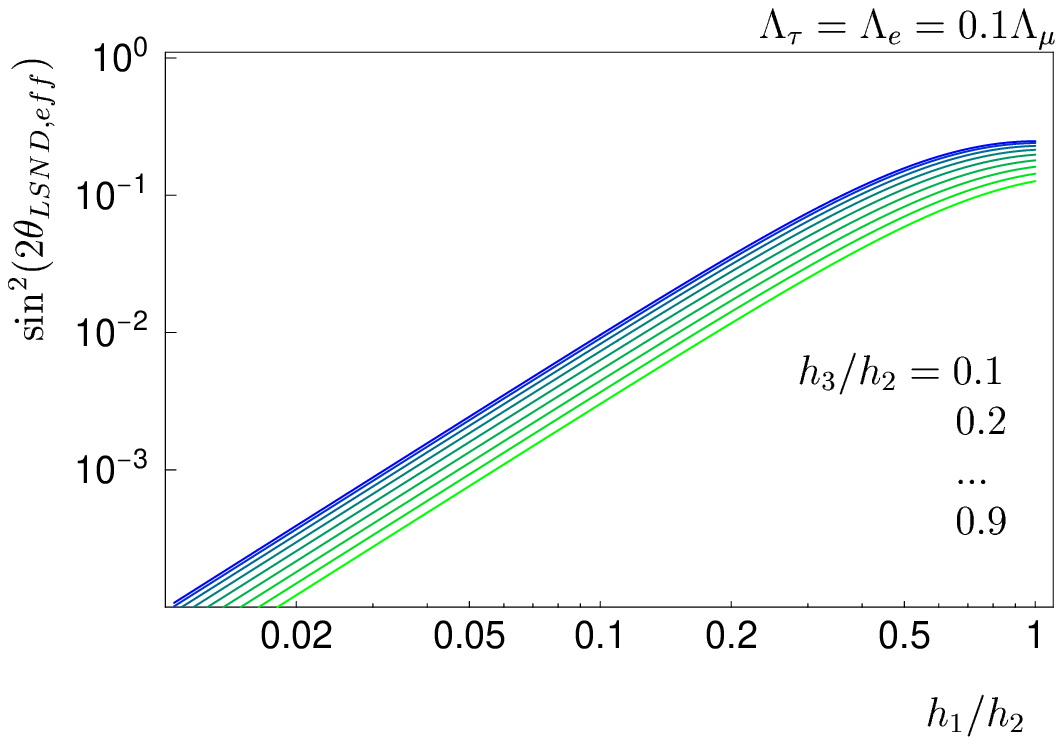,height=20.0cm,width=15.0cm}}}
\end{picture}
\caption[]{
  The ``effective'' LSND angle for the neutrino spectrum of
  \fig{4nuspec} as a function of $h_1/h_2$ for different choices of
  $h_3/h_2$. Obviously to explain the LSND result $h_1 \ll h_2$ is
  required.}
\label{lsndeff}
\end{figure}

Since LSND is sensitive only to the largest mass gap in the spectrum,
this corresponds simply to a re-interpretation of the effective mixing
angle. In \fig{lsndeff} we show $\sin^2(2\theta_{LSND,eff})$ as a
function of $h_1/h_2$ for different values of $h_3/h_2$. For 
$\sin^2(2\theta_{LSND,eff}) \sim 10^{-2}-10^{-3}$, $h_1/h_2 \sim 0.02-0.1$ 
is needed.

Slightly more complicated is the conversion probability for the 
atmospheric neutrinos, given as
\ba{cpatm}\nn
P_{\nu_{\mu} \to   \nu_x}^{atm} &=& 
P_{\nu_{\mu} \to   \nu_{\tau}}^{atm} + 
P_{\nu_{\mu} \to   \nu_s}^{atm} \\ \nn
& = & 4 \sin^2(\Delta m^2_{LSND} \frac{L}{E}) 
\{ |(U_{\mu 1}U_{\tau 1}+U_{\mu 2} U_{\tau 2}) 
(U_{\mu 3} U_{\tau 3}+ U_{\mu 4} U_{\tau 4})| 
+ | {\rm \tau \to  s} | \}\\
& + &  4 \sin^2(\Delta m^2_{atm} \frac{L}{E}) \{ | U_{\mu 3} U_{\tau 3} 
U_{\mu 4} U_{\tau 4}| + | {\rm \tau \to  s} | \}
\ea

The first term proportional to $\sin^2(\Delta m^2_{LSND} \frac{L}{E})$
will be averaged over in the atmospheric neutrino data (except for the
smallest $L$ and largest $E$) and should be visible in the data as a
finite ``offset'' proportional to $2 \{ |(U_{\mu 1}U_{\tau 1}+U_{\mu
  2} U_{\tau 2}) (U_{\mu 3} U_{\tau 3}+ U_{\mu 4} U_{\tau 4})| + |
{\rm \tau \to s} | \}$. The term $4 \{ | U_{\mu 3} U_{\tau 3} U_{\mu
  4} U_{\tau 4}| + | {\rm \tau \to s} | \}$ defines the effective
atmospheric neutrino mixing angle in our model.

\begin{figure}
\setlength{\unitlength}{1mm}
\begin{picture}(0,70)
\put(-25,-85)
{\mbox{\epsfig{figure=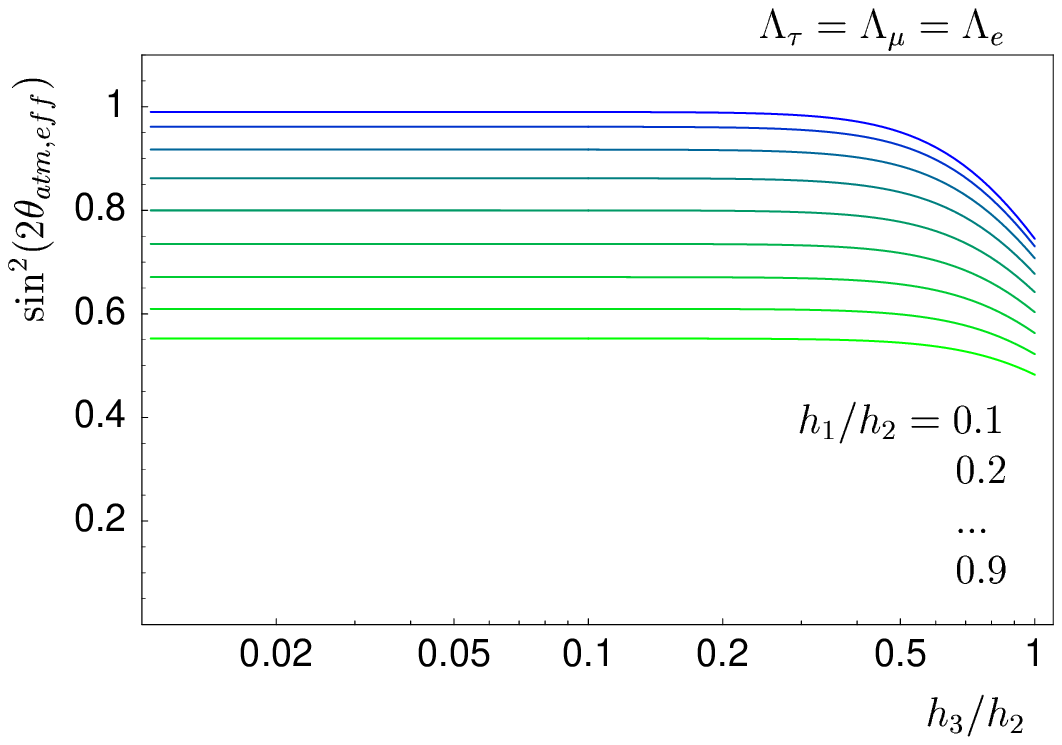,height=20.0cm,width=15.0cm}}}
\put(50,-85)
{\mbox{\epsfig{figure=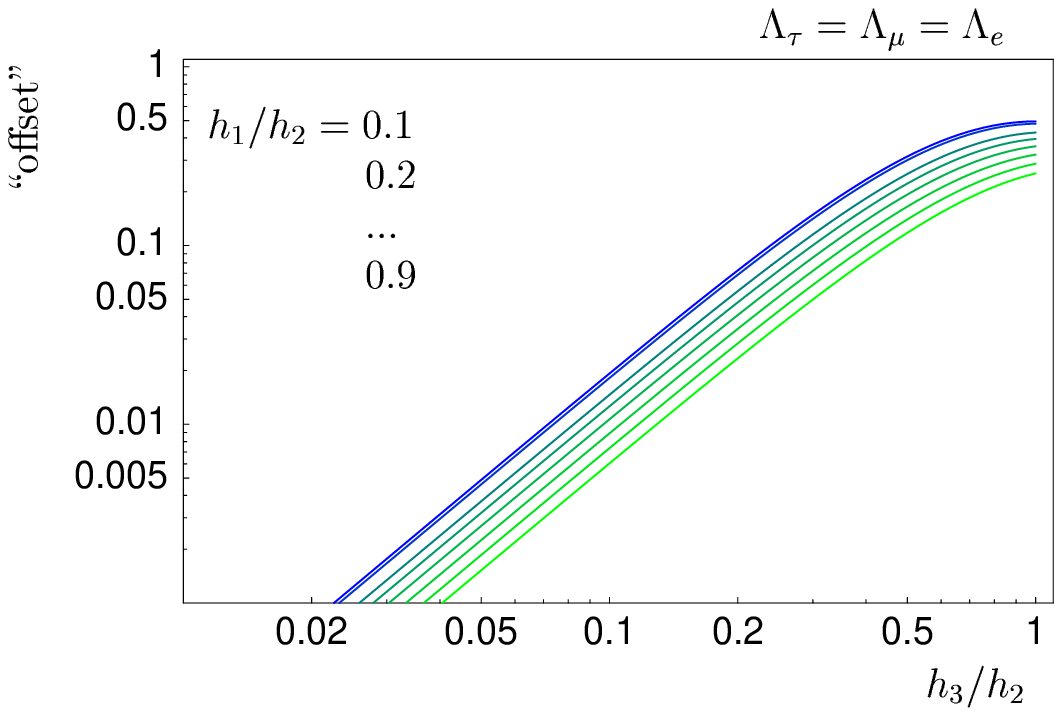,height=20.0cm,width=15.0cm}}}
\end{picture}
\caption[]{The ``effective'' atmospheric neutrino angle (left) and 
  ``offset'' (right) as a function of $h_3/h_2$ for different
  $h_1/h_2$ choices. }
\label{atmeff}
\end{figure}

In \fig{atmeff} we give the ``effective'' atmospheric neutrino angle
(left) and the corresponding ``offset'' (right) as a function of
$h_3/h_2$ for different choices of $h_1/h_2$. The size of this offset
depends mainly on $h_3/h_2$. Note that both reactor and LSND data
require $h_1 \ll h_2$.

Note that in the (unrealistically) extreme case of $h_1$, $h_3 \to 0$
we would have pure (two generation) $\nu_{\mu} \to \nu_s$ oscillations
in our model. However, relatively large values of $h_3$ are not
excluded by the data, implying that the atmospheric neutrino
oscillations in our model are described in general by a mixture of
$\nu_{\mu} \to \nu_s$ and $\nu_{\mu} \to \nu_{\tau}$.  Note also that
pure $\nu_{\mu} \to \nu_{\tau}$ oscillations are not allowed in our
model, since the quasi-Dirac pair between which the oscillation takes
place necessarily involves the sterile state. We have estimated that
at least 50 \% of the oscillation probability must be due to
$\nu_{\mu} \to \nu_s$ in our model.

It has been stated that present atmospheric neutrino data rule out a
pure $\nu_{\mu} \to \nu_s$ conversion \cite{ichepsolat00}. However, we
find it premature to confidently rule out this possibility at this
stage. Indeed, from a global fit in which uncertainties are treated
conservatively \cite{Fornengo:2000sr}, this exclusion does not yet
emerge, although the pure sterile conversion is indeed disfavored.
However, in four-neutrino models like the present, the atmospheric
conversions are certainly not pure $\nu_{\mu} \to \nu_s$ in general,
as seen from eq.  \rf{cpatm}.  As a result, even if ones rules out
pure $\nu_{\mu} \to \nu_s$ conversion one does not rule out the model
itself. However it is definitely clear that it will be tested by more
refined data yet to come. The bound on the admixture of $\nu_{\mu} \to
\nu_s$ can be inferred from the atmospheric data as suggested in
\cite{Yasuda:2000de}. Moreover, the ``offset'' predicted to exist in
our model, if the atmospheric oscillations are not pure $\nu_{\mu} \to
\nu_s$, might be testable by the K2K experiment.

\begin{figure}
\setlength{\unitlength}{1mm}
\begin{picture}(0,70)
\put(-25,-85)
{\mbox{\epsfig{figure=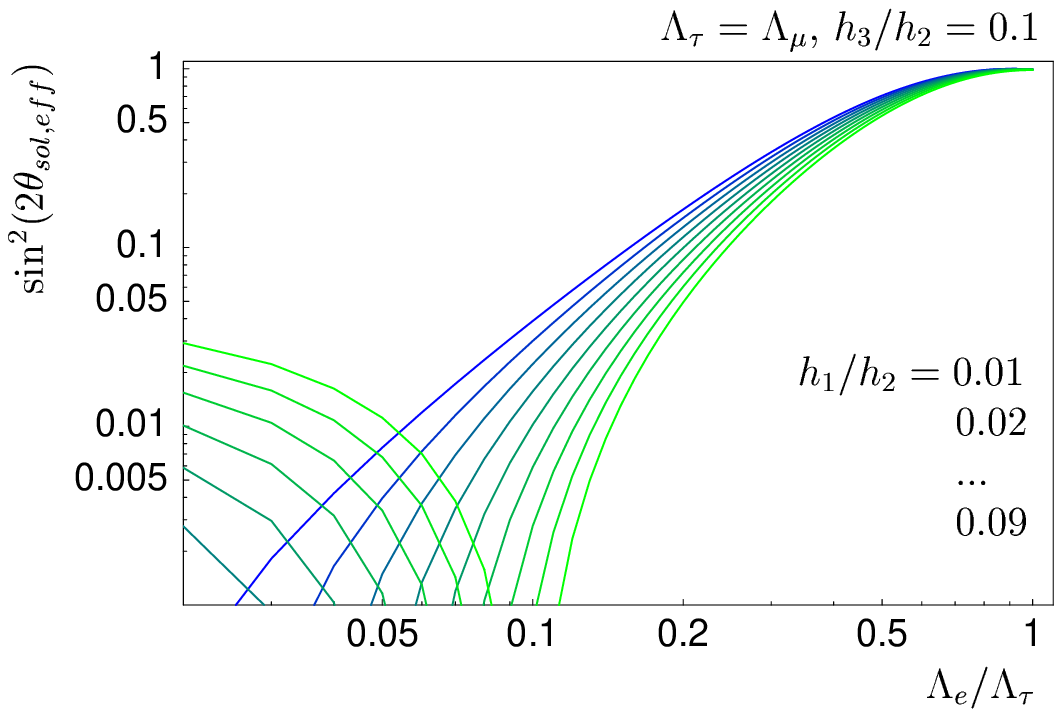,height=20.0cm,width=15.0cm}}}
\put(50,-85)
{\mbox{\epsfig{figure=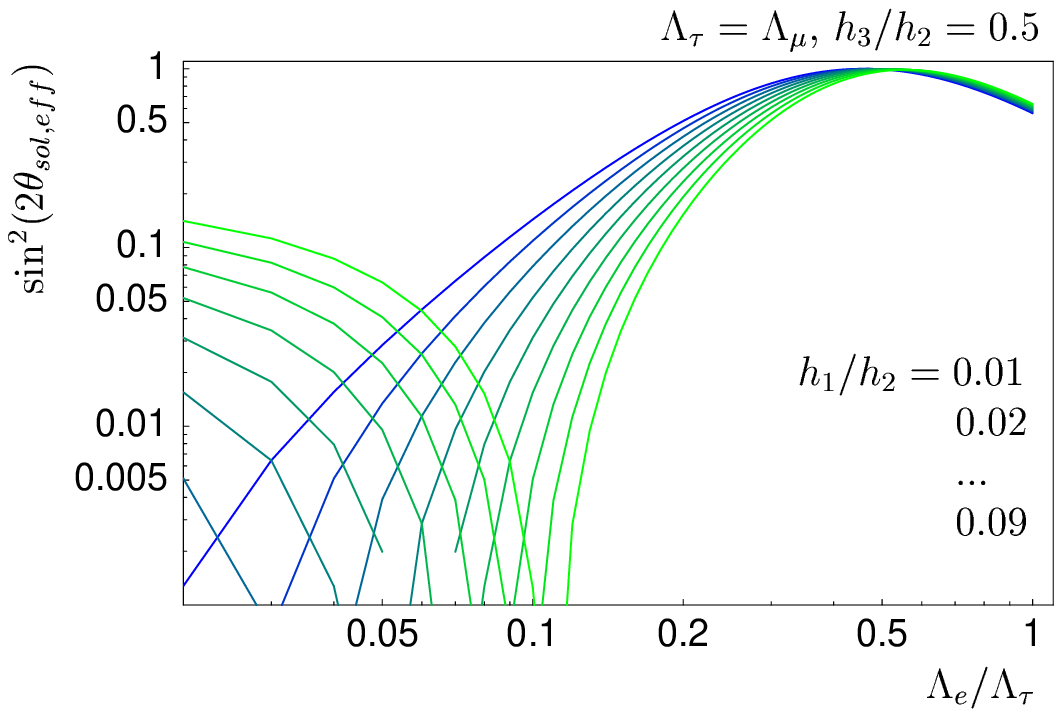,height=20.0cm,width=15.0cm}}}
\end{picture}
\caption[]{The ``effective'' solar angle as a function of 
$\Lambda_e/\Lambda_{\tau}$ for two different choices of $h_3/h_2$. }
\label{soleff}
\end{figure}

To close the discussion on neutrino angles, in \fig{soleff} we display
the effective solar angle as a function of $\Lambda_e/\Lambda_{\tau}$
for two different choices of other parameters.  Due to the flatness of
the recoil electron neutrino spectrum indicated by the most recent
solar neutrino data~\cite{ichepsolat00}, the solutions to the solar
neutrino problem which are now preferred by the data involve large
mixing~\cite{update00,MSW00}.  Large neutrino mixing angles require
$\Lambda_e \simeq \Lambda_{\tau}$ within a factor of $2-3$.

\section{Neutralino decay}

Let us finally briefly discuss the decay of the lightest of the 4
heavy states in our neutrino-neutralino spectrum. We will call it for
simplicity the lightest neutralino. As is
well-known with the presence of the bilinear (lepton number violating)
terms in the superpotential the lightest neutralino is no longer
stable \cite{Romao:1996mg,Bartl:2000yh} and decays through lepton
number violating modes. In the bilinear R-parity violating model it
has been shown that this decay occurs with sizeable branching ratio
into visible states \cite{Hirsch:2000ef} inside typical detector sizes
despite the small neutrino masses required by current neutrino data.
Moreover, these decays might be used to obtain information on neutrino
angles \cite{porod2000}.

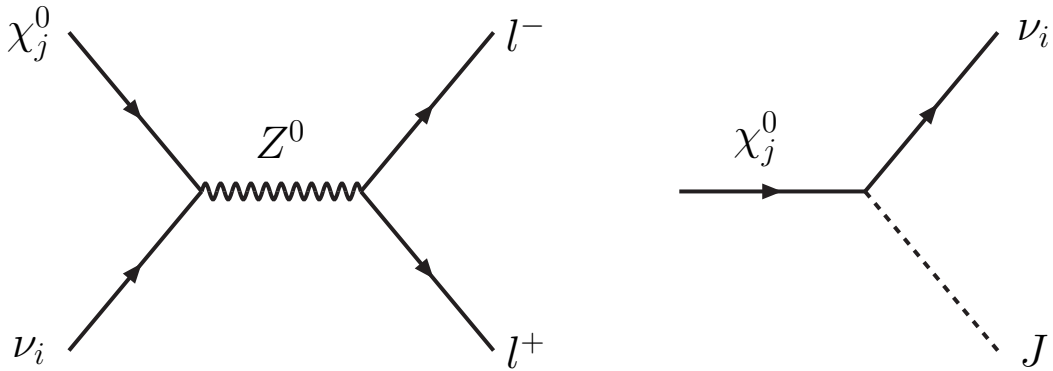
\begin{figure}
\begin{picture}(340,185)(0,0)
\SetWidth{1.5}
\ArrowLine(20,0)(70,60)
\ArrowLine(20,120)(70,60)
\Photon(70,60)(130,60){3}{10.5}
\ArrowLine(130,60)(180,0)
\ArrowLine(130,60)(180,120)
\Text(100,80)[]{{\Large $Z^0$}}
\Text(5,120)[]{{\Large $\chi^0_j$}}
\Text(5,0)[]{{\Large $\nu_i$}}
\Text(193,120)[]{{\Large $l^-$}}
\Text(193,0)[]{{\Large $l^+$}}
\ArrowLine(250,60)(320,60)
\ArrowLine(320,60)(370,120)
\DashLine(320,60)(370,0){3}
\Text(280,80)[]{{\Large $\chi^0_j$}}
\Text(385,0)[]{{\Large $J$}}
\Text(385,120)[]{{\Large $\nu_i$}}
\end{picture}
\caption[]{Two examples of Feynman graphs leading to the decay of the 
lightest neutralino in our model. To the left, neutral current interaction 
leading to visible final states. To the right, invisible decay into 
a Majoron and a neutrino.}
\label{decay}
\end{figure}

Since in our present model R-parity violation is required in order to
provide the solar and atmospheric neutrino mass splittings, the
neutralino will decay as well. However, two features distinguish the
current model from the simpler bilinear model. First, the spontaneous
violation of R-parity (which is used as a seed for the generation of
the bilinear terms) implies the existence of a Majoron.  The
neutralino then also can decay via the invisible mode $\chi^0 \to J
\nu$~\cite{Romao:1996mg}, see \fig{decay}.

However, due to the smallness of the (Dirac)
neutrino coupling this decay mode will be very much suppressed when
compared to the visible \rp violating decays. A very crude 
order-of-magnitude estimate for the two graphs shown in \fig{decay} 
gives for the neutral current graph 

\be{decayvis}
\Gamma^{vis} \sim \frac{g^2}{16 \pi} \big(\frac{|\Lambda|}{M_W^2}\big)^2 
m_{\chi^0} \sim 10^{-16} \hskip2mm {\rm GeV}
\ee
while for the Majoron decay one expects: 
\be{decayinvis}
\Gamma^{invis} \sim \frac{|h^{\nu}|^2}{16 \pi} 
m_{\chi^0} \sim 10^{-22} \hskip2mm {\rm GeV}
\ee

Thus, even though a Majoron exists in the present model, sufficiently 
large branching ratios to visible states in the decay of $\chi^0$ exist 
to be searched for at accelerators.

Second and more important, however, in the present model the solar
neutrino problem is solved by $\nu_e \to \nu_{\tau}$ oscillations
(with possibly some admixture of $\nu_e \to \nu_{\mu}$).  If the solar
neutrinos indeed require large mixing, as currently favoured by the
data, $\Lambda_e \simeq \Lambda_{\tau}$ is needed in our model to fit
the data. This would then lead to the prediction that in the decays of
the neutralino to semi-leptonic final states a comparable number of
electrons and taus should exist. This prediction might be used to
distinguish the present model from the bilinear R-parity violating
model, since the latter requires that electrons in the semi-leptonic
states should be very much suppressed \cite{Hirsch:2000ef,porod2000}
reflecting the fact that in the bilinear model $\Lambda_e \ll
\Lambda_{\mu} \simeq \Lambda_{\tau}$ is required.

\section{Conclusion}

We have discussed a very economical model of neutrino mass which,
despite its simplicity is in principle able to fit all neutrino data
including LSND.
The model is minimal in the sense that we only introduce one additional 
\21 singlet superfield into the MSSM superpotential. 
The LSND scale can be explained by the fermion present in this
superfield, while its scalar partner can trigger the spontaneous
violation of R-parity, through a nonzero right-handed sneutrino vacuum
expectation value, which in turn splits the Dirac neutrino into a
quasi-Dirac pair (atmospheric neutrino scale), providing also mass to
one of the low--lying neutrinos at the solar neutrino scale.
The smallness of the overall neutrino mass (LSND scale) can be
ascribed to the volume suppression factor associated with the
compactified dimensions. 
Thus the model can naturally explain all neutrino oscillation data,
making four predictions for the neutrino oscillation parameters.
We stress that this model is testable in the near future: one of its
necessary ingredients is that atmospheric neutrino conversions must
include at least some $\nu_{\mu} \to \nu_s$ oscillations. Thus future
atmospheric neutrino measurements will provide a crucial test of this
model.
However, if tides turn, and sterile neutrinos in the future are not
disfavored by Superkamiokande, another test of the model could be done
at accelerators. The model predicts that the lightest neutralino
decays inside the detector with sizeable branching ratios into visible
states.

\newpage
\bigskip
\centerline{\bf Acknowledgement}

This work was supported by DGICYT grant PB98-0693 and by the TMR
contract ERBFMRX-CT96-0090. M.H. acknowledges support from the
European Union's Marie-Curie program under grant No ERBFMBICT983000.

\end{document}